\documentclass[twocolumn,secnumarabic,amssymb, nofootinbib, nobibnotes, aps, prd, reprint,superscriptaddress]{revtex4-1}

\usepackage{bm}            
\usepackage[cmex10]{amsmath}
\usepackage{amssymb}
\usepackage{mathrsfs}
\usepackage{mathtools}
\usepackage{feynmp-auto}
\usepackage[british]{babel}
\usepackage[titletoc,toc,title]{appendix}
\usepackage{scrextend}

\usepackage{color}

\begin{document}
\title{Simulations of relativistic-quantum plasmas using real-time lattice scalar QED}
\author         {Yuan Shi}
\email          {yshi@pppl.gov}
\affiliation    {Department of Astrophysical Sciences, Princeton University, Princeton, NJ 08544 USA}
\affiliation    {Princeton Plasma Physics Laboratory, Princeton University, Princeton, NJ 08543 USA}

\author         {Jianyuan Xiao}
\affiliation    {School of Nuclear Science and Technology and Department of Modern Physics, University of Science and Technology of China, Hefei, Anhui 230026, China}

\author         {Hong Qin}
\affiliation    {Department of Astrophysical Sciences, Princeton University, Princeton, NJ 08544 USA}
\affiliation    {Princeton Plasma Physics Laboratory, Princeton University, Princeton, NJ 08543 USA}
\affiliation    {School of Nuclear Science and Technology and Department of Modern Physics, University of Science and Technology of China, Hefei, Anhui 230026, China}

\author         {Nathaniel J. Fisch}
\affiliation    {Department of Astrophysical Sciences, Princeton University, Princeton, NJ 08544 USA}
\affiliation    {Princeton Plasma Physics Laboratory, Princeton University, Princeton, NJ 08543 USA}

\date           {\today}

\begin{abstract}
Real-time lattice quantum electrodynamics (QED)	provides a unique tool for simulating plasmas in the strong-field regime, where collective plasma scales are not well-separated from relativistic-quantum scales. As a toy model, we study scalar QED, which describes self-consistent  interactions between charged bosons and electromagnetic fields. To solve this model on a computer, we first discretize the scalar-QED action on a lattice, in a way that respects geometric structures of exterior calculus and U(1)-gauge symmetry. The lattice scalar QED can then be solved, in the classical-statistics regime, by advancing an ensemble of statistically equivalent initial conditions in time, using classical field equations obtained by extremizing the discrete action.
To demonstrate the capability of our numerical scheme, we apply it to two example problems. The first example is the propagation of linear waves, where we recover analytic wave dispersion relations using numerical spectrum. The second example is an intense laser interacting with a 1D plasma slab, where we demonstrate natural transition from wakefield acceleration to pair production when the wave amplitude exceeds the Schwinger threshold. 
Our real-time lattice scheme is fully explicit and respects local conservation laws, making it reliable for long-time dynamics. The algorithm is readily parallelized using domain decomposition, and the ensemble may be computed using quantum parallelism in the future. 
\end{abstract}

\maketitle
\setlength{\parskip}{0pt}
\section{Introduction}
Lattice QED, a scheme usually used to study vacuum quantum electrodynamics, can also be used to simulate plasmas. By adding dynamical background fields, we extend lattice QED into a valuable tool for plasma physics, especially when plasmas are dense or when fields are strong. Under these extreme conditions where collective QED effects are important, the commonly adopted plasma kinetic model, which arises as the geometrical optics approximation of the relativistic-quantum world \cite{Ruiz15}, is no longer sufficient. An example where QED effects are important is the production of electron-positron pairs when intense lasers interact with plasma targets \cite{Liang98,Gahn00,Liang15,Sarri15}. To describe such phenomena in semiclassical framework, source terms must be inserted into quantum kinetic or fluid equations \cite{Berezhiani92,Kluger98,Schmidt98,Roberts02,Hebenstreit10}, which can then be solved by numeric integration \cite{Hebenstreit08,Hebenstreit09} or QED-PIC simulations \cite{Duclous11,Nerush11,Ridgers12}. However, prefabricated source terms take little account of the interplay between coexisting processes \cite{Schutzhold08}, nor can they describe quantum interference, through which the created pairs may be entangled. Therefore, while semiclassical approximations may be applicable for long-wavelength lasers, large errors are expected when fields, such as those of x-ray lasers, evolve on scales comparable to intrinsic QED scales. Moreover, in semiclassical treatments, there is no obvious way to subtract both energy and momentum from fields once pairs are produced. Although errors may be tolerable when fields dominate particles, disrespecting energy-momentum conservation will likely have nonphysical consequences, after a large number of pairs are generated. 

To model plasmas where QED processes have no clear scale separation from classical processes, a faithful description can only be provided on the relativistic-quantum level. While lattice simulations may be unfamiliar tools for plasma physics, they have been used extensively in quantum chromodynamics (QCD) to describe the strong interaction \cite{Wilson74} and quark-gluon plasmas \cite{Bass99,Satz00}. In conventional lattice-QCD simulations, quantum correlation functions are computed using numerical path integrals, from which observables are extracted as coefficients of scaling laws \cite{Creutz80}. This scheme can be analytically continued to imaginary time to describe statistical systems in thermal equilibrium \cite{Yagi05}. For out-of-equilibrium systems, real-time simulations can be carried out using the Schwinger-Keldysh time contours \cite{Schwinger61,Keldysh65}. The above formulations, based on numerical path integrals, are capable of capturing genuine quantum loop effects, but are numerically expensive. Fortunately, the computational cost can be dramatically reduced when the occupation numbers of quantum states are high and when the coupling is weak. This is precisely the case for plasma physics, where a large number of particles are present, and the coupling coefficient $e\approx0.3$ is small. In this classical-statistic regime, tree-level effects dominate loop effects \cite{Aarts02,Mueller04,Jeon05,Berges07,Berges14}, and the quantum system can be adequately described by time-advancing the classical field equations with an ensemble of statistically equivalent initial conditions \cite{Aarts99,Gelis13,Polkovnikov03,Borsanyi09}. Based on this approach, lattice spinor-QED simulations have been carried out to demonstrate production of fermion pairs from the vacuum by a prescribed external electric field in one spatial dimension \cite{Hebenstreit13,Hebenstreit13String,Kasper14}. However, the role of background plasmas during pair production has not been investigated. By incorporating a nonperturbative amount of background particle fields, we turn real-time lattice simulations into numerical tools useful for plasma physics.

In this paper, we demonstrate plasma effects during pair productions using the scalar-QED model, where the electromagnetic (\textit{EM}) fields evolve in a self-consistent manner, instead of being imposed from boundary and initial conditions. Using the scalar-QED model, we avoid the fermion doubling problem when discretizing the Dirac field \cite{Chandrasekharan04} and focus on unambiguous plasma contributions. The scalar-QED model governs interactions between \textit{EM} fields and spin-0 charged bosons, such as charged pions or Cooper pairs. Although laboratory plasmas are typically made of spin-1/2 charge fermions, classical plasma physics takes no account of particle spin-statistics at all. Therefore, to demonstrate that lattice simulations are useful for plasma physics, it is sufficient to study scalar QED, which has been used to describe laser-plasma interaction in relativistic-quantum regime \cite{Eliasson11,Shi16}. In the classical-statistics regime, scalar QED is governed by Klein-Gordon-Maxwell (KGM) equations
\begin{eqnarray}
\label{eq:KGMp}
(D_{\zeta,\mu} D_\zeta^\mu+m_\zeta^2)\phi_\zeta&=&0, \\
\label{eq:KGMA}
\partial_\mu F^{\mu\nu}&=&j^{\nu},
\end{eqnarray}
where we have used the natural units $\hbar=c=\epsilon_0=1$. In the above equations, $\phi_\zeta$ is the complex scalar field, describing spin-0 bosons of species $\zeta$, whose charge is $q_\zeta$ and mass is $m_\zeta$. The real-valued 1-form $A_\mu$ is the gauge field, describing spin-1 bosons, and $F_{\mu\nu}=\partial_\mu A_\nu-\partial_\nu A_\mu$ is the field strength tensor. Charged bosons couple to the gauge field through the covariant derivative $D_{\zeta,\mu}=\partial_\mu-iq_\zeta A_\mu$, and the gauge field couples with charged fields through the gauge-invariant current density
\begin{equation}
\label{eq:j}
j^\mu=\sum_\zeta\frac{q_\zeta}{i}\big[\bar{\phi}_\zeta(D_\zeta^\mu\phi_\zeta)-\text{c.c.}\big],
\end{equation}
where $\bar{\phi}_\zeta$ denotes complex conjugation of $\phi_\zeta$. By the famous Klein paradox \cite{Klein29}, the charged scalar field $\phi_\zeta$ cannot be interpreted as the probability amplitude of a single particle. A more appropriate interpretation is that the classical field $\phi_\zeta$ is intrinsically a many-particle field, which can be represented as $\phi_\zeta(x)=\int\sqrt{V}\Phi_\zeta(x,x_2,x_3,\dots)$, where $\Phi_\zeta(x,x_2,x_3,\dots)$ is the symmetrized many-body wave function, and the integration is carried out on the many-body configurations space \cite{Shi16}. Regardless of the interpretation, we can solve the KGM equations as coupled partial differential equations, whose solutions model the tree-level behavior of charged bosons interacting with \textit{EM} fields.

This paper is organized as follows. In Sec.~\ref{sec:algorithm}, we develop a variational algorithm for solving the KGM equations. 
In Sec.~\ref{sec:examples}, we apply this algorithm to two example problems in plasma physics. The first example is the propagation of linear waves, where we compare numerical spectra with analytical dispersion relations. The second example is wakefield acceleration and pair production, when intense lasers interact with a 1D plasma slab. 
Conclusion and discussion are given in Sec.~\ref{sec:conclusion}. In Appendix~\ref{appendix:A}, we discuss local conservation laws underlying our algorithm. In Appendix~\ref{appendix:B}, we summarize an explicit numerical scheme using the Lorenz gauge condition.

\section{\label{sec:algorithm}Variational algorithm}
In the continuum, the KGM equations can be derived from the action $S=\int d^4x\mathcal{L}$, where the Lorentz-invariant and U(1)-gauge-invariant scalar-QED Lagrangian 
\begin{equation}
\label{eq:L}
\mathcal{L}=(\overline{D_{\mu}\phi})(D^{\mu}\phi)-m^2\bar{\phi}\phi-\frac{1}{4}F_{\mu\nu}F^{\mu\nu}.
\end{equation}
Here we have omitted the species subscript $\zeta$, and the summation of charged species is implied. By Noether's theorem, the U(1)-gauge symmetry of the action
\begin{equation}
\phi\rightarrow\phi e^{iq\alpha}, \hspace{5pt} A_\mu\rightarrow A_\mu+\partial_\mu\alpha,
\end{equation}
implies charge conservation $\partial_\mu j^\mu=0$, where the current density $j^{\mu}$ is given by Eq.~(\ref{eq:j}). Similarly, by the Lorentz symmetry, energy and momentum are also conserved $\partial_\mu\mathcal{T}^{\mu\nu}=0$, where the gauge invariant stress-energy tensor 
\begin{eqnarray}
\label{eq:T}
\nonumber
\mathcal{T}^{\mu\nu}&=&(\overline{D^{\mu}\phi})(D^{\nu}\phi)+(D^{\mu}\phi)(\overline{D^{\nu}\phi})\\
&+&F^{\mu\sigma}F_{\sigma}^{\phantom{\mu}\nu}-g^{\mu\nu}\mathcal{L}.
\end{eqnarray}
Here, $g^{\mu\nu}$ is the Minkowski metric with characteristics $(+,-,-,-)$. This scalar-QED theory, omitting the $\phi^4$ self-coupling, is the underlying model of our algorithm on the discrete spacetime lattice. In fact, a variational algorithm for solving the KGM equations has already been developed in the numerical analysis community \cite{Christiansen11}, which shows superior charge conservation property when gauge symmetry is respected. In this paper, we rederive the variational algorithm in arbitrary gauge, using local energy conservation to justify the choice of Yee-type action \cite{Yee66} over Wilson-type action \cite{Wilson74}, and emphasize on the application of such algorithm to plasma physics.

\subsection{\label{sec:discrete}Discretization of fields and action}
\begin{figure}[t]
	\renewcommand{\figurename}{FIG.}
	\includegraphics[angle=0,width=6.5cm]{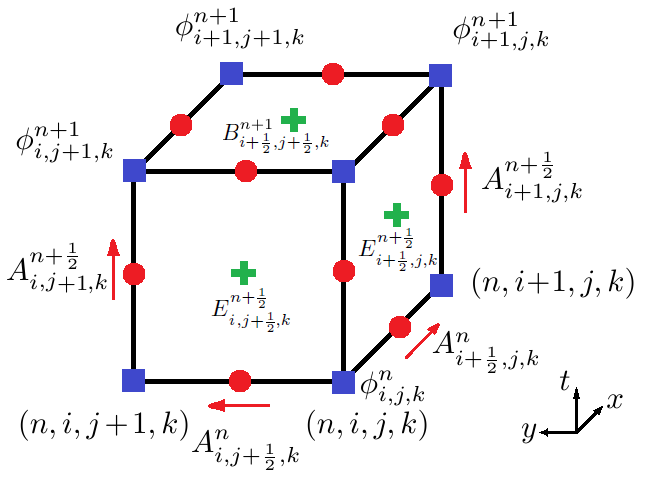}
	\caption{Discretization of the $txy$-submanifold of spacetime. The discrete function $\phi_v$ lives on the vertexes (blue squares). For example, $\phi^{n}_{i,j,k}=\phi(t_n,x_i,y_j,z_k)$ lives on the vertex $(n,i,j,k)$. 
	The discrete 1-form $A_e$ lives along edges (red circles). For example, the $t$-component $A_{i+1,j,k}^{n+1/2}=A^{0}(t_n+\Delta t/2,x_{i+1},y_j,z_k)$ lives along the edge connecting vertexes $(n,i+1,j,k)$ and $(n+1,i+1,j,k)$, and the $x$-component $A_{i+1/2,j,k}^{n}=-A^{1}(t_n,x_i+\Delta x/2,y_j,z_k)$ lives along the edge connecting vertexes $(n,i,j,k)$ and $(n,i+1,j,k)$. 
	The discrete 2-form $F_f$ lives on faces (green crosses). For example, electric field $E_{i+1/2,j,k}^{n+1/2}=E^x(t_n+\Delta t/2,x_i+\Delta x/2,y_j,z_k)$ lives on the time-like face spanned by vertexes $(n,i,j,k), (n+1,i,j,k), (n+1,i+1,j,k)$ and $(n,i+1,j,k)$; magnetic field $B_{i+1/2,j+1/2,k}^{n+1}=B^z(t_{n+1},x_i+\Delta x/2,y_j+\Delta y/2,z_k)$ lives on the space-like face spanned by vertexes $(n+1,i,j,k), (n+1,i,j+1,k), (n+1,i+1,j+1,k)$ and $(n+1,i+1,j,k)$.
	}
	\label{fig:FieldDisc}
\end{figure}

To solve the continuous system numerically, let us discretize the spacetime manifold using a rectangular lattice. Then the scalar field $\phi$, namely, a function on the spacetime manifold, naturally lives on the vertexes of the discrete manifold
\begin{equation}
\phi^{n}_{i,j,k}:=\phi(t_n,x_i,y_j,z_k),
\end{equation} 
where $(t_n,x_i,y_j,z_k)$ is the coordinate of the vertex. In comparison, the gauge 1-form $A=A_\mu dx^{\mu}$ naturally lives along the edges of the discrete spacetime manifold. For example, the $t$- and $x$-components
\begin{eqnarray}
A_{i,j,k}^{n+\frac{1}{2}}&:=&+A^{0}(t_n+\frac{\Delta t}{2},x_i,y_j,z_k),\\
A_{i+\frac{1}{2},j,k}^{n}&:=&-A^{1}(t_n,x_i+\frac{\Delta x}{2},y_j,z_k),
\end{eqnarray}
where $\Delta t=t_{n+1}-t_n$ and $\Delta x=x_{i+1}-x_i$. The minus sign comes from the Minkowski metric $g_{\mu\nu}$, which lower the index $A_\mu=g_{\mu\nu}A^{\nu}$. In the above discretization, a half-integer index indicates which edge does the field resides along. For example, $A_{i,j,k}^{n+1/2}$ resides along the edge connecting vertices $(t_n,x_i,y_j,z_k)$ and $(t_{n+1},x_i,y_j,z_k)$, and is therefore the $A_0$ component of $A$. 
Notice that since $A$ is a 1-form living along edges, only one of its four indices can take half-integer values, while the other three indices must take integer values. Moreover, to each edge of the lattice, the discrete 1-form only assigns the component of $A$ that is parallel to this edge (Fig.~\ref{fig:FieldDisc}), while other components of $A$ are not assigned.  

Now that we have discretized the fields, the gauge-covariant derivatives can be computed using the Wilson's lines \cite{Wilson74}. Since the gauge-covariant derivatives are 1-forms, they also lives along edges when discretized. For example, the $t$- and $x$-components of the pull-back gauge covariant derivatives are
\begin{eqnarray}
\label{eq:D0}
(D_0^<\phi)_{i,j,k}^{n+\frac{1}{2}}\!&=&\!\frac{1}{\Delta t}\Big(\phi^{n+1}_{i,j,k}e^{-iq\Delta tA_{i,j,k}^{n+\frac{1}{2}}}-\phi^{n}_{i,j,k}\Big),\\
\label{eq:D1}
(D_1^<\phi)_{i+\frac{1}{2},j,k}^{n}\!&=&\!\frac{1}{\Delta x}\Big(\phi^{n}_{i+1,j,k}\!e^{-iq\Delta xA_{i+\frac{1}{2},j,k}^{n}}\!-\!\phi^{n}_{i,j,k}\Big).\hspace{10pt}
\end{eqnarray}
These pull-back covariant derivatives transform under the discrete U(1)-gauge symmetry [Eqs.~(\ref{eq:U1phi})-(\ref{eq:GaugeAi})] as $\phi_{i,j,k}^n$ [Eq.~(\ref{eq:Cov})]. Analogously, one can define push-forward covariant derivatives, which we shall not use in this paper. The gauge field $A$ serves as the 1-form defining the connection on the U(1)-bundle, which enables parallel transport $\phi$ on the spacetime manifold.

To compute the field strength tensor $F_{\mu\nu}$, notice that $F=dA$ is the curvature 2-form and hence lives on faces of the lattice upon discretization. For example, the time-like component $F_{01}=E^1$ is the electric field in the $x$-direction, which can be computed by
\begin{equation}
E_{i+\frac{1}{2},j,k}^{n+\frac{1}{2}}\!=\!\frac{A_{i+\frac{1}{2},j,k}^{n+1}\!-\!A_{i+\frac{1}{2},j,k}^{n}}{\Delta t}
\!-\!\frac{A_{i+1,j,k}^{n+\frac{1}{2}}\!-\!A_{i,j,k}^{n+\frac{1}{2}}}{\Delta x}.
\end{equation}
This component lives on the time-like face spanned by four vertices $(n,i,j,k)$, $(n,i+1,j,k)$, $(n+1,i+1,j,k)$, and $(n+1,i,j,k)$. 
Analogously, we can compute the space-like components of $F$. For example, $F_{12}=-B^3$ is the magnetic field in the $z$-direction
\begin{eqnarray}
\nonumber
-B_{i+\frac{1}{2},j+\frac{1}{2},k}^{n}&=&\frac{1}{\Delta x}\bigg(A_{i+1,j+\frac{1}{2},k}^{n}\!-\!A_{i,j+\frac{1}{2},k}^{n}\bigg)\\
&-&\frac{1}{\Delta y}\bigg(A_{i+\frac{1}{2},j+1,k}^{n}\!-\!A_{i+\frac{1}{2},j,k}^{n}\bigg).
\end{eqnarray}
This \textit{z}-component of the magnetic field lives on the space-like face spanned by four vertices $(n,i,j,k)$, $(n,i+1,j,k)$, $(n,i+1,j+1,k)$ and $(n,i,j+1,k)$. 
Notice that the sign of the discretized $F$ is determined by the orientation of the face. 
Since the above discretization respects geometric structures of exterior calculus, the Bianchi identities, namely, $\nabla\cdot\mathbf{B}=0$ and the Faraday's law, are automatically and exactly satisfied (Appendix~\ref{appendix:A}). 

Using the discrete gauge-covariant derivatives and the discrete field strength, the action can be discretized by
\begin{eqnarray}
\label{eq:Sd}
S_d=\sum_{c}\Delta V L_d[\phi_v,A_e],
\end{eqnarray}
where $\phi_v$ and $A_e$ are the discrete fields. Here the subscript $v$ denotes vertexes, and $e$ denotes edges. In the discrete action, $\Delta V$ is the volume 4-form, and the summation runs over all cells of the lattice. In each unit cell, the discrete Lagrangian function 
\begin{eqnarray}
\label{eq:Ld}
L_d=(\overline{D_\mu\phi})_e(D^\mu\phi)_e-m^2\bar{\phi}_{v}\phi_v+\frac{1}{2}(E_f^2-B_f^2),
\end{eqnarray}
where summations over unique $e$, $v$, and $f$ are implied. Here, the subscript $f$ denotes faces. Notice that in contrast to what is typically done in lattice gauge theories, here we have directly used the discrete field strength $F_{\mu\nu}F^{\mu\nu}$, instead of the Wilsonian plaquettes $L_w\propto\text{Re}[1-\exp(ieF_{\mu\nu}\Delta_\mu\Delta_\nu)]$. We made this choice so that a discrete version of the local energy conservation law for \textit{EM} fields is exactly satisfied when the coupling goes to zero (Appendix \ref{appendix:A}) .

\subsection{\label{sec:EOM}Equations of motion for discrete fields}
\begin{figure}[t]
	\renewcommand{\figurename}{FIG.}
	\includegraphics[angle=0,width=6cm]{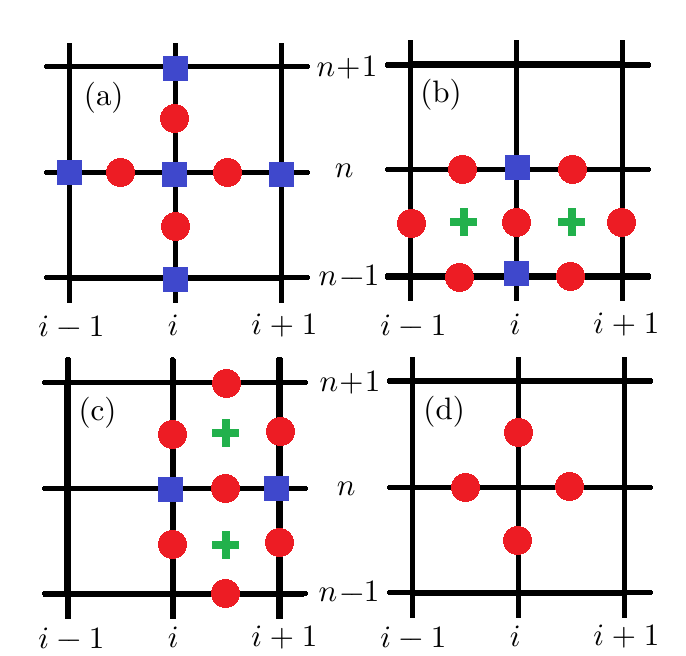}
	\caption{Coupling pattern of $\phi_v$ (blue squares), $A_e$ (red circles) and $F_f$ (green crosses) in the $tx$-submanifold. (a) The discretized KG equation [Eq.~(\ref{eq:phid})] couples $\phi_{v}$ with its nearest neighbors though $A_e$. (b) The discretized Gauss' law [Eq.~(\ref{eq:Ed})] couples $E_{i-1/2}$ and $E_{i+1/2}$ through $\phi_v$ on the shared vertexes. (c) The Maxwell-Amp\`ere's law [Eq.~(\ref{eq:Bd})] couples $E^{n+1/2}$ to $E^{n-1/2}$ through $\phi_v$ on the shared vertexes, and through $B^n$ (not depicted here) who shares the common space-like edge. (d) The Lorenz gauge condition couples $A_e$ that shares the same vertex. }
	\label{fig:Coupling}
\end{figure}

Having discretized the action, the classical equation of motion (EOM) for the discrete field $\phi_v$ can be obtained by extremizing $S_d$. Taking variation with $\bar{\phi}_v$ and set $\delta S_d/\delta \bar{\phi}_{v}=0$, a discrete version of equation of Eq.~(\ref{eq:KGMp}) is
\begin{eqnarray}
\label{eq:phid}
&&\frac{1}{\Delta t^2}\bigg(\phi_{s}^{n+1}e^{-iq\Delta tA_{s}^{n+\frac{1}{2}}}-2\phi_{s}^{n} +\phi_{s}^{n-1}e^{iq\Delta tA_{s}^{n-\frac{1}{2}}}\bigg)\\
\nonumber
&=&\frac{1}{\Delta_l^2}\bigg(\phi_{s+l}^{n}e^{-iq\Delta_lA_{s+\frac{l}{2}}^{n}}\!-\!2\phi_{s}^{n} +\phi_{s-l}^{n}e^{iq\Delta_lA_{s-\frac{l}{2}}^{n}}\bigg)\!-\!m^2\phi_{s}^{n},
\end{eqnarray}
where the time index is explicit, the vertex-centered spatial index is abbreviated as $s=(i,j,k)$, and summations in $l=i,j,k$ directions are implied.
By taking variation with $\phi_v$, we can obtain the EOM for $\bar{\phi}_v$, which is the complex conjugation of the above equation. The finite difference equation (\ref{eq:phid}) is centered around vertexes, and couples $\phi_{v}$ with its eight nearest neighbors though $A_e$, as illustrated by Fig.~\ref{fig:Coupling}(a) in the $tx$-submanifold.

To find the equation for the electric field, take variation of $S_d$ with respect to the time-like component $A_{s}^{n+1/2}$. Setting $\delta S_d/\delta A_{s}^{n+1/2}=0$, we obtain a discrete version of the Gauss' law, centered along time-like edges
\begin{eqnarray}
\label{eq:Ed}
\frac{1}{\Delta_l}\Big(E_{s+\frac{l}{2}}^{n+\frac{1}{2}}-E_{s-\frac{l}{2}}^{n+\frac{1}{2}}\Big)=J^{n+1/2}_{s}.
\end{eqnarray}
The charge density 1-form $J^{n+1/2}_{s}$ is the hodge dual of the charge density 3-form $j_0=\star j^0$, discretized by 
\begin{equation}
\label{eq:charge}
J^{n+1/2}_{s}=\frac{iq}{\Delta t}\Big(\bar{\phi}_{s}^{n+1}e^{iq\Delta tA_{s}^{n+\frac{1}{2}}}\phi_{s}^{n}-\text{c.c.}\Big).
\end{equation}
When there are multiple charged species, the right-hand side (RHS) should sum over charge densities of all species. In Fig.~\ref{fig:Coupling}(b), we illustrate the coupling pattern of the above finite difference equation.

To find the equations involving components of the magnetic field, take variation of $S_d$ with respect to the space-like components $A_{s+l/2}^{n}$. For example, by setting $\delta S_d/\delta A_{i+1/2,j,k}^{n}=0$, we obtain an equation advancing the \textit{x}-component of the electric field in time by
\begin{eqnarray}
\label{eq:Bd}
\frac{E_{s+\frac{i}{2}}^{n+\frac{1}{2}}-E_{s+\frac{i}{2}}^{n-\frac{1}{2}}}{\Delta t}=\epsilon_{ijk}\frac{B_{r-\frac{k}{2}}^{n}-B_{r-\frac{k}{2}-j}^{n}}{\Delta_j}+J^{n}_{s+\frac{i}{2}}.
\end{eqnarray}
Here, $r=(i+1/2, j+1/2, k+1/2)$ is the abbreviated index for the body center, $\epsilon_{ijk}$ is the Levi-Civita symbol, and summations over repeated indexes are implied. The current density 1-form $J^{n}_{s+i/2}$ is the hodge dual of the current density 3-form $j_i=\star j^i$. The hodge dual gives rise to a negative sign, so that the \textit{x}-component of the current density $-j^x$ is discretized by 
\begin{equation}
\label{eq:current}
J^{n}_{s+\frac{l}{2}}=\frac{iq}{\Delta_l}\Big(\bar{\phi}_{s+l}^{n}e^{iq\Delta_l A_{s+\frac{l}{2}}^{n}}\phi_{s}^{n}-\text{c.c.}\Big).
\end{equation}
The finite difference equation (\ref{eq:Bd}) is the discrete version of the Maxwell-Amp\`ere's law  $\partial_tE^x=\partial_yB^z-\partial_zB^y-j^x$ centered around space-like edges, whose coupling pattern is illustrated in Fig.~\ref{fig:Coupling}(c). When computing $j_x$ on the RHS, summation over charged species is implied.

In order to advance the above finite difference equations in time, we need to fix a gauge to eliminate the extra degree of freedom. Since the discrete action $S_d$ is U(1)-gauge invariant (Appendix \ref{appendix:A}), we can choose any gauge. For example, one convenient choice is the Lorenz gauge  $\partial_\mu A^\mu=0$. When discretized, the Lorenz gauge condition allows time advance $A_{s}^{n-1/2}\rightarrow A_{s}^{n+1/2}$ in a very simple way [Eq.~(\ref{eq:An}), Fig.~\ref{fig:Coupling}(d)]. Another convenient choice is the temporal gauge $A^0=0$. When discretized, $A_{s}^{n+1/2}$ remains zero on all time-like edges.

Having obtained discrete equations and fixed the gauge, an explicit time advance scheme can be constructed (Appendix~\ref{appendix:B}). We first initialize the simulation by giving values of $\phi_s^n$ at both $n=0$ and $n=1$ for every spatial lattice points $s$. This is necessary because the KG equation [Eq.~(\ref{eq:KGMp})] is a second order partial differential equation and therefore needs two initial conditions. Similarly, we need to give initial values of $A_e$ at $n=0$ and $n=1/2$. Second, we use the discrete Gauss' Law to calculate $A_e$ at $n=1$ by solving a system of linear equations [Eq.~(\ref{eq:Gauss}), Fig.~\ref{fig:Coupling}b]. Notice that the continuous version of this equation can be written as $\partial_t\nabla\cdot\mathbf{A}=-\nabla^2A^0-\rho$, where the RHS is known. Because the unknowns on the left-hand side (LHS) involve only first order spatial derivative, the discrete Gauss' law couples less number of points than the discrete Poisson's equation, and is therefore easier to solve. Third, we use the Lorenz gauge condition to advance $(A^{n-1/2},A^{n})\rightarrow A^{n+1/2}$ [Eq.~(\ref{eq:Lorenz}), Fig.~\ref{fig:Coupling}d]. Fourth, we use the discrete KG equation to calculate $\phi^{n+1}$ in terms of $\phi^{n}$ and $\phi^{n-1}$ [Eq.~(\ref{eq:phin}), Fig.~\ref{fig:Coupling}a]. This involves exponentiation of $A^n$ and $A^{n+1/2}$, whose values are already known at this point. Simultaneously, we can compute $A^{n+1}$ in terms of $A_e$ at previous time steps [Eq.~(\ref{eq:An}), Fig.~\ref{fig:Coupling}c], using the discrete Maxwell-Amp\`ere's law. Notice that the discrete Gauss' Law is preserved during time advance, which is a consequence of the discrete local charge conservation law [Eq.~(\ref{eq:Jconserv})]. Finally, having computed both $\phi_v$ and $A_e$ at $t=n+1$, we can move forward in the time loop by updating $n\rightarrow n+1$, with proper boundary conditions supplied. In similar fashion, explicit time advance schemes can be constructed when other gauge conditions are used. 

\begin{figure}[t]
	\renewcommand{\figurename}{FIG.}
	\includegraphics[angle=0,width=8.5cm]{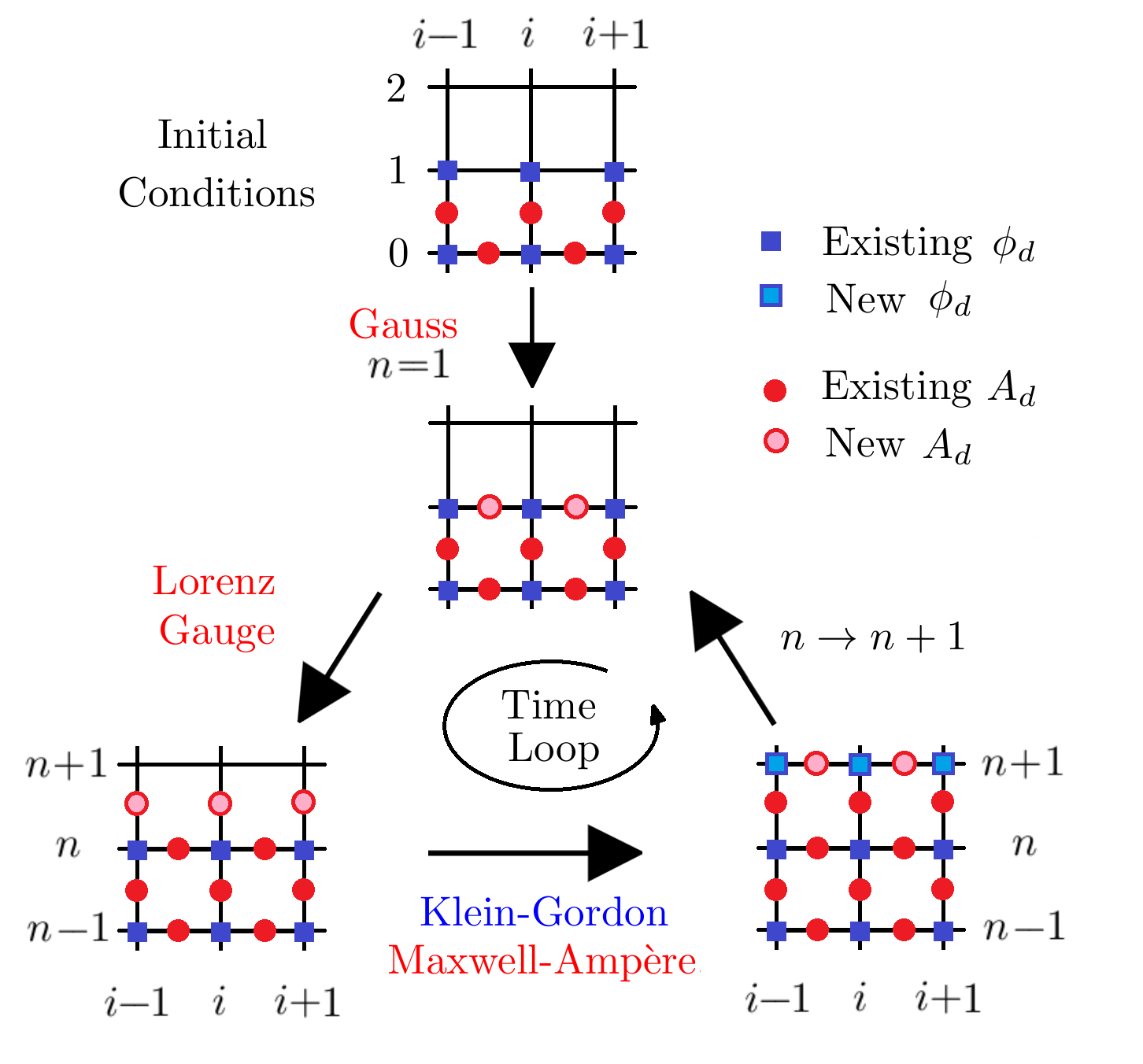}
	\caption{Time evolution scheme for discrete KGM equations using the Lorenz gauge.  As initial conditions, the values of $\phi_v(n=0)$ and $\phi_v(n=1)$ are given (blue squares), so are $A_e(n=0)$ and $A_e(n=1/2)$ (red circles). Then the Gauss' Law [Eq.~(\ref{eq:Ed})] is used to calculate $A_e(n=1)$. On entering the time loop, the first step is to calculate $A^{n+1/2}$ using the Lorenz gauge condition. The second step is to use the KG equation [Eq.~(\ref{eq:phid})] to calculate $\phi^{n+1}$, and independently, use the Maxwell-Amp\`ere's law [Eq.~(\ref{eq:Bd})] to calculate $A^{n+1}$. The time loop is advanced by $n\rightarrow n+1$ and repeat. }
	\label{fig:Schematics}
\end{figure}

\section{\label{sec:examples}Numerical examples}
In this section, we demonstrate our numerical scheme using two examples. The first example is the propagation of linear waves, and the second example is laser-plasma interaction in one spatial dimensional (1D).

\subsection{\label{sec:wave}Linear wave spectra}
To test our code implementation, we compare numerical spectra with analytical linear wave dispersion relations \cite{Hines78,Kowalenko85,Eliasson11,Shi16}. For small-amplitude waves, the dispersion relation constrains the wave frequency $\omega$ as function of the wave vector $\mathbf{k}$. In unmagnetized cold scalar-QED plasma, the tree-level dispersion relation of the transverse \textit{EM} wave is 
\begin{equation}
\label{eq:EMD}
\omega^2=\omega_p^2+\mathbf{k}^2,
\end{equation}
where $\omega_{p}^2=\sum_\zeta\omega_{p\zeta}^2$ is the total plasma frequency, and $\omega_{p\zeta}^2=q_\zeta^2n_{\zeta0}/m_\zeta$ is the plasma frequency of individual charged species $\zeta$. 
The other eigenmode is the longitudinal electrostatic wave, whose dispersion relation is
\begin{equation}
\label{eq:ESD}
1+\chi_p=0,
\end{equation}
where the cold plasma susceptibility
\begin{equation}
\chi_p=\sum_{\zeta}\frac{\omega_{p\zeta}^{2}(\mathbf{k}^2-\omega^2+4m_{\zeta}^2)}{(\omega^2-\mathbf{k}^2)^2-4m_{\zeta}^2\omega^2}.
\end{equation} 
The dispersion relation of electrostatic wave contains three branches. The gapless branch is the acoustic wave, the low frequency branch is the Langmuir mode, and the high frequency branch is the pair mode. While acoustic mode and Langmuir mode exist in classical plasmas, the pair mode only exists in relativistic-quantum plasmas \cite{Fuda82}. The pair mode can be excited when gamma photons $(\omega>2m)$ inelastically scatter in high density plasmas, creating longitudinal oscillations in which virtual pairs are created and annihilated to carry the wave quanta.

 \begin{figure}[b]
 	\renewcommand{\figurename}{FIG.}
 	\includegraphics[angle=0,width=0.5\textwidth]{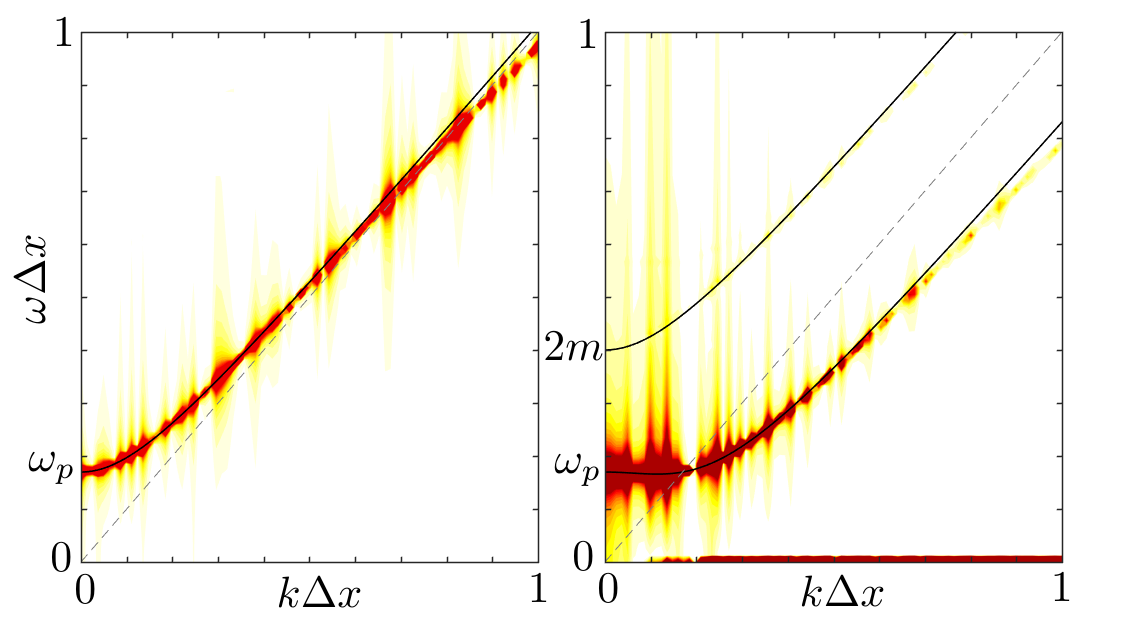}
 	\caption{Power spectra (color) of the transverse electric field $E_y$ (a) and the longitudinal electric field $E_x$ (b) are well-traced by tree-level dispersion relations (black lines) up to the grid resolution. The power spectra are averaged over an ensemble of 100 simulations with statistically equivalent initial conditions. In these simulations, immobile ion background is homogeneous. The charge $q=0.3$, such that the fine structure constant $q^2/4\pi\approx1/137$ is physical. The unperturbed background plasma density is extremely high, such that the plasma frequency $\omega_p=0.85m$ can be shown on the same scale as $m$. 
 	The resolution $m\Delta x=0.04$ and $m\Delta t=0.02$. The number of spatial grid point is $L=512$, and the total number of time steps, including the initial conditions, is $T=1024$. The dashed gray lines is the light cone.}
 	\label{fig:Dispersion}
 \end{figure}
 
We compute the numerical spectra in a single species plasma, in which immobile ions serve as homogeneous neutralizing background. To initialize the simulation so that a broad spectrum of linear waves are excited, the initial values of $A_e$ are given using small amplitude white noise with mean $\mu(A_e)=0$ and standard deviation $\sigma(A_e)=10^{-4}m$. Assuming the charged field is initially free, then its initial conditions can be given using the free field expansion $\phi(x)=\int d^3\mathbf{p} [a_\mathbf{p}\exp(-ipx)+b^{\dagger}_\mathbf{p}\exp(ipx)](2\pi)^{-3} (2E_\mathbf{p})^{-1/2}$,
where $px=E_\mathbf{p}t-\mathbf{p}\cdot\mathbf{x}$ is Minkowski inner product, and $E_\mathbf{p}=\sqrt{\mathbf{p}^2+m^2}$ is the relativistic energy corresponding to momentum $\mathbf{p}$. 
From the above expansion, the momentum space distribution functions for particles and antiparticles are $f_a(\mathbf{p})=a_\mathbf{p}^\dagger a_\mathbf{p}$ and the $f_b(\mathbf{p})=b_\mathbf{p}^\dagger b_\mathbf{p}$, respectively. Consider the simple example where the plasma is initially homogeneous and constituted of cold particles, namely, $f_a(\mathbf{p})=n_0\delta^{(3)}(\mathbf{p})$ and $f_b(\mathbf{p})=0$, where $n_0$ is the background plasma density. Then, the free charged field $\phi(x)=\sqrt{n_0/2m}\exp(-imt+i\alpha)$, where $\alpha$ is some random phase. When discretized, this free field corresponds to the initial conditions $\phi_v^0=\sqrt{n_0/2m}\exp(i\alpha)$ and $\phi_v^1=\phi_s^0\exp(-im\Delta t)$. An ensemble of statistically equivalent initial conditions can then be constructed by randomly sample the phase $\alpha$ and the gauge field.

After advancing the initial conditions in time using periodic boundary conditions, numerical spectra can be read out from simulations by taking discrete Fourier transforms of electric field components. Since the unmagnetized plasma is isotropic, it is sufficient to read out the dispersion relation in the \textit{tx}-submanifold. In this submanifold, the spectra of either $E_y$ or $E_z$ correspond to the dispersion relation of transverse \textit{EM} modes, and the spectrum of $E_x$ corresponds to the dispersion relation of longitudinal electrostatic modes. 
The ensemble-averaged power spectrum of $E_y$ [Fig.~\ref{fig:Dispersion}(a)] is indistinguishable from that of $E_z$, and is well-traced by the analytical dispersion relation (black line) of the transverse \textit{EM} wave [Eq.~(\ref{eq:EMD})], until $k\Delta x\sim 1$ where the spatial resolution is no longer sufficient.  
Similarly, the ensemble-averaged power spectrum of $E_x$ [Fig.~\ref{fig:Dispersion}(b)] is localized near three bands, corresponding to the cold acoustic mode, the Langmuir mode and the pair mode [Eq.~(\ref{eq:ESD})]. That the analytical dispersion relations are recovered by numerical spectra indicates that our solutions faithfully capture the propagation of linear waves up to the grid resolution.

\subsection{\label{sec:pair}Laser-plasma interaction}
Having verified our code implementation, let us study laser-plasma interaction as another example, which can no longer be described self-consistently under the classical framework once the laser wavelength becomes too short and the field strength becomes too large. Before discussing our simulations in this relativistic-quantum regime, it is helpful to recall what happens in the classical regime \cite{Kruer88}. Classically, when the plasma slab is under-dense, namely when the laser frequency $\omega>\omega_p$, much of the laser will travel through the plasma slab, with some reflection and inverse Bremsstrahlung absorption. In an initially quiescent slab, the laser will propagate uneventfully, if its frequency stays away from the two-plasmon-decay resonance, and its intensity is not strong enough to grow instabilities within the pulse duration. Beyond nonlinear wave instabilities, when the laser field becomes relativistically strong, namely when $a=qE/m\omega\gtrsim1$, the ponderomotive force of a short laser pulse can expels a significant fraction of plasma electrons and form wakefield \cite{Pukhov02}. The wakefield can then accelerate particles, generating energetic beams of particles and radiations trailing the laser pulse. When the beams are energetic enough, they may produce gamma photons through synchrotron radiation or Bremsstrahlung. The virtual gamma photons may then decay into electron-positron pairs through the trident process \cite{Bjorken67}. Alternatively, the on-shell gamma photons may interact with ion potentials and produce pairs through the Breit-Wheeler process \cite{Beit34}. Finally, when the laser field becomes even stronger, namely when $qE/m^2\gtrsim 1$, pairs may also be produced directly through the Schwinger process \cite{Schwinger51}. 

Many aspects of laser-plasma interaction can be studied using our new numerical tool. Here, to validate that our code can capture genuine relativistic-quantum effects, we select parameters in our 1D simulations to demonstrate transition from wakefield acceleration to Schwinger pair production as we increase the laser intensity. Notice that in 1D, the phase space is highly constrained. 
Using periodic boundary conditions in directions transverse to laser propagation, Schwinger pair production by laser fields is suppressed. This is because when transverse fields try to pull $e^-/e^+$ pairs apart, their wave functions are enforced to be the same by the periodic boundary condition, which prevents pairs from emerging out of vacuum fluctuations. Therefore, in 1D simulations, Schwinger pair production requires longitudinal field $E_x$. To generate $E_x$ beyond the Schwinger field $E_c=m^2/q$ through wakefield, the plasma density must be extremely high. Heuristically, to produce on-shell pairs, the critical electric field needs to separate the pair by Compton wavelength $1/m$ within the Compton time $T\sim\pi/m$, namely, $qE_xT^2/m\gtrsim1/m$. In the wave-breaking regime, $E_x\simeq a m\omega_p/q$, so the inequality requires that the plasma density be high enough such that the plasma frequency $\omega_p/m\gtrsim1/a\pi^2$. In reality, at those densities, we should treat the electron Fermi degeneracy to capture the full physical effects. However, simulating instead a high-density bosonic plasma is just a toy model that tests our code, with the density picked so high that we can already see laser Schwinger pair production in 1D simulations.

With this basic understanding of how laser pair production happen in 1D, we choose parameters to suppress the trident and Breit-Wheeler processes, by treating ions as immobile homogeneous neutralizing background, so that there is no spiky ion potentials from which energetic ``electrons" and gamma photons can scatter. The smooth ion background provides an electrostatic potential that initially confines the ``electrons". We initialize the charged boson wave function according to $\phi(x)=\sqrt{n_0(x)/2m}\exp(-imt)$, where $n_0(x)$ is the background ion density with a plateau of width $L\approx100/m$ and Gaussian off-ramps with $\sigma=20/m$. For density of the bosonic plasma to be high enough to enable pair production, we take $n_0=m^3$ so that the plasma frequency $\omega_p=0.3m$ is enormous. The above wave function is a linear superposition of many eigenstates of the system. In our simulations, we let the wave function evolve to statistically stationary states through phase mixing, before we start to draw samples at random time intervals. The sampled wave functions are then used as initial conditions for $\phi_v$, which are combined with initial values $A_e$ of a Gaussian pulse to construct an ensemble. The linearly-polarized Gaussian pulse is initialized in the vacuum region with zero carrier phase $A_y\propto\exp(-\xi^2/2\tau^2)\cos\omega\xi$, where $\xi=x-t$ and $\tau=20/m$. For the laser to be able to transmit the high-density plasma slab, we use a gamma-ray laser with frequency $\omega_0=0.7m$, for which semiclassical treatments are far from valid. The laser envelope is slowly varying ($\omega_0\tau=14$), and has full width at half maximum about twice the plasma skin depth. When intense laser pulse propagates, it can excite plasma waves, from which the laser can be Raman scattered.

\begin{figure}[t]
	\renewcommand{\figurename}{FIG.}
	\includegraphics[angle=0,width=0.5\textwidth]{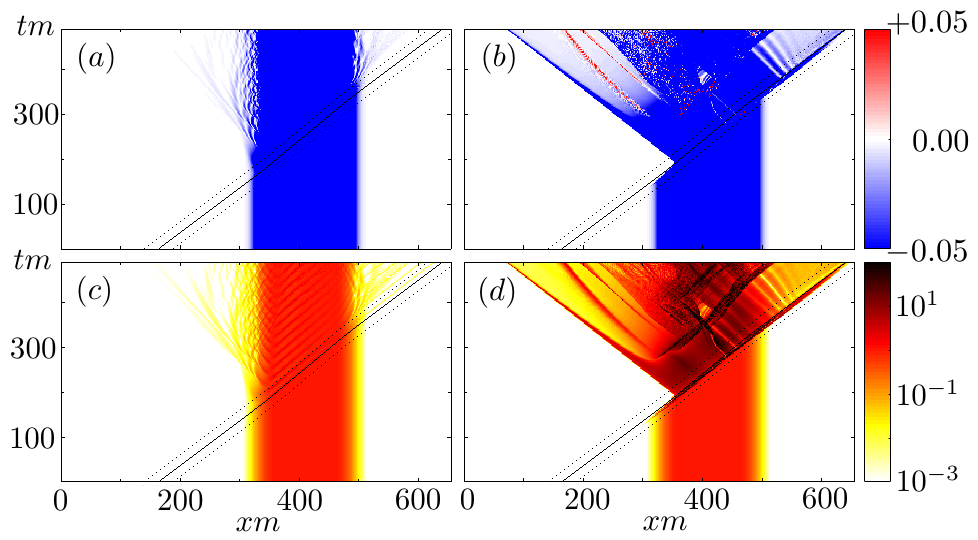}
	\caption{Charge density (a, b) and energy density (c, d) of the $\phi$ field. 
	When the gamma-ray laser ($\omega_0=0.7m$) is relativistic $(a\approx1)$, but not strong enough to produce Schwinger pairs $(E_x\approx0.3E_c)$, ``electrons" are expelled by the laser ponderomotive force, accelerated by the wakefield, and splashed from the plasma boundaries (a, c). 
	When the laser field exceeds the Schwinger threshold $(a\approx16, E_x\approx5E_c)$, copious pairs are produced when laser interacts with plasma waves (b, d). The spin-0 ``electrons" are initially confined by a smooth immobile neutralizing background, with a density plateau $n_0=m^3$ and a Gaussian off-ramp $\sigma =20/m$. The trajectories of the pulse center (black lines) and the pulse half widths (dashed lines) are well traced by geometric optics. Both the charge density (normalized by $em^3$) and the energy density (normalized by $m^4$) are averaged over an ensemble of size 200. The resolutions are such that $m\Delta x=0.04$ and $m\Delta t=0.005$.}
	\label{fig:part}
\end{figure}

\begin{figure}[b]
	\renewcommand{\figurename}{FIG.}
	\includegraphics[angle=0,width=0.5\textwidth]{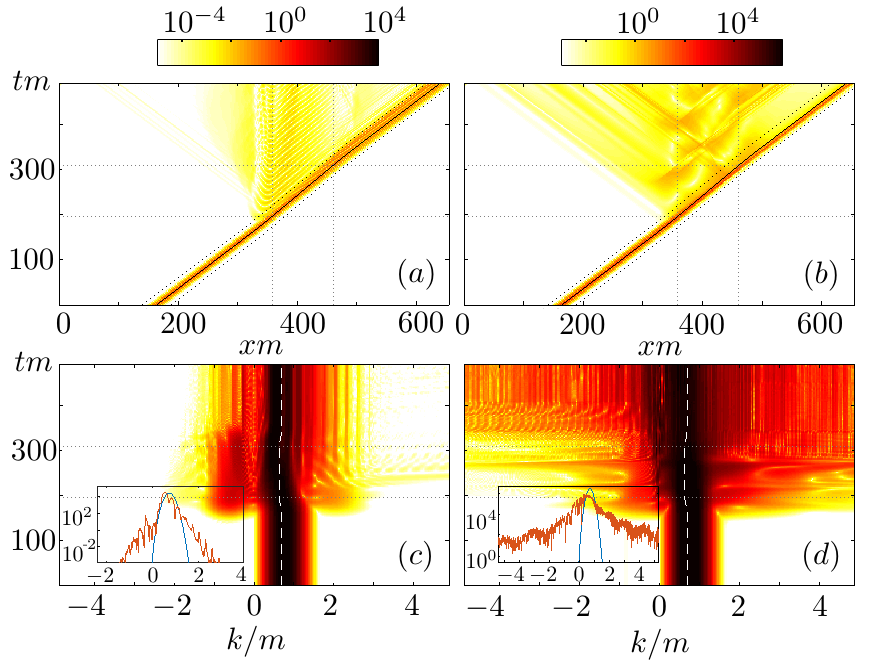}
	\caption{Total energy density of \textit{EM} fields (a, b), and the power spectral density of its transverse components (c, d). The inserts show the initial (blue) and final (red) spectra of \textit{EM} waves. 
	When $a\approx1$ ($E_x\approx0.3E_c$) is below the Schwinger field, the laser excites plasma waves and is Raman scattered (a, c). The time evolution of the main pulse is well-traced by geometric optics (dashed lines). 
	When the laser field $a\approx16$ ($E_x\approx5E_c$) is above the Schwinger field, a noticeable amount of energy is lost due to pair production (b), and the $k$-spectrum is substantially broadened (d). 
	The field energy density is normalized by the Schwinger field $E_c^2$, and are averaged over an ensemble of size 200. The resolutions are such that $m\Delta x=0.04$ and $m\Delta t=0.005$. The dotted gray lines mark where the geometric-optics trajectory of the pulse center crosses the plasma plateau boundaries.}
	\label{fig:field}
\end{figure}

With the above setup, the laser pulse simply travels through the plasma with some refraction and reflections when the laser field is weak ($a\ll 1$). More interesting phenomena happen when the laser field becomes strong. For example, when $a\approx1$ is relativistically strong but the resulting $E_x\approx0.3E_c$ is below the Schwinger field, our simulation recovers what happens in classical plasmas \cite{McKinstrie96,Naumova04,Geyko09}. First, let us look at what happens to charged particles. After the laser enters the plasma, beams of ``electrons" are formed in the forward direction by both ponderomotive snow-plow and laser wakefield acceleration. At the same time, some ``electrons" are splashed in the backward direction from strongly-driven plasma boundaries (Fig.~\ref{fig:part}a, c). 
Next, for the laser pulse, its center (solid black lines) and half widths (dotted black lines) are well-traced by geometric optics in the $xt$-space (Fig.~\ref{fig:field}a), as well as in the $kt$-space (Fig.~\ref{fig:field}c, dashed white line), because the background plasma is smooth on the laser wavelength scale. Beyond geometric optics, as the laser travels through the plasma slab, ponderomotive expulsion of ``electrons" cause the laser pulse to adiabatically loose a small amount of energy in the form of frequency redshift $\omega<\omega_0$ (Figs.~\ref{fig:field}a,c and \ref{fig:total}b). In addition, the laser excites plasma waves, from which the laser is Raman-scattered in both forward and backward directions. In the insert of Fig.~\ref{fig:field}c, the final spectrum (red) shows distinctive Raman scattering peaks at $\omega+n\omega_p$ up to $n=8$, and second harmonics peaks $2\omega$ and $2\omega+\omega_p$ in the forward direction. In the backward direction, peaks at $\omega-\omega_p, \omega, \omega+\omega_p$ and $2\omega$ can also be identified unambiguously.

When we increase the laser field beyond the Schwinger threshold ($a_c=m/\omega$). For example, when $a\approx16$ ($E_x\approx5E_c$), a large amount of $e^-/e^+$ pairs are produced (Figs.~\ref{fig:part}b, d). A very small fraction of pairs are produced and trapped in the laser wakefield, forming low-luminosity ``electron" (negative charge density, blue) and ``positron" (positive charge density, red) beams that leave the plasma slab from its right boundary. In contrast, a much larger fraction of pairs are produced when the backscattered \textit{EM} wave, whose intensity is near the Schwinger threshold (Fig.~\ref{fig:field}b), interacts with forward-propagating plasma waves. ``Positrons" produced in this way form high-luminosity collimated beams, leaving the plasma slab from its left boundary. Apart from these beams, many ``positrons" never manage to leave the plasma slab. These trapped ``positrons" have large probabilities to annihilate with ``electrons" in the highly constrained 1D phase space. 
Due to pair creation and particle acceleration, the laser initially looses a significant amount of energy, until pair creation and annihilation roughly balance (Figs.~\ref{fig:field}b,c and \ref{fig:total}b). At that point, the $k$-spectrum of the laser is substantially broadened (Fig.~\ref{fig:field}d). Such a spectral broadening is expected from general wave action considerations \cite{Wilks88,Dodin10}, which predict frequency up-shift due to pair creation, and frequency down-shift due to pair annihilation and plasma expulsion. In the insert of Fig.~\ref{fig:field}d, the final \textit{EM} wave spectrum (red) shows distinctive annihilation bumps near integer multiples of ``electron" rest mass. These annihilation peaks are very broad since ``electrons" and ``positrons" annihilate with large kinetic energy. Finally, notice that no pair is produced when the laser travels through the vacuum region, which is expected in 1D. It is remarkable that very rich physics can already be captured by simply solving the classical field equations with proper initial conditions.  

To extract observables from simulations, the charge density (Figs.~\ref{fig:part}a, b) is computed using Eq.~(\ref{eq:charge}), which includes no contribution from background ions. Therefore, negative charge (blue) indicates ``electron" density in excess of ``positron" density, whereas positive charge (red) indicates the contrary. The energy density of the charged field (Figs.~\ref{fig:part}c, d) and the \textit{EM} fields (Figs.~\ref{fig:field}a, b) are computed using Eqs.~(\ref{eq:Hphi}) and (\ref{eq:HA}), respectively. To compute $k$-spectra of \textit{EM} waves (Figs.~\ref{fig:field}c, d), notice that a monochromatic \textit{EM} wave satisfies $k_xE_y=\omega B_z$. 
Upon discretization, this relation remains exactly satisfied if we take $k_x=\sin(k\Delta x)/\Delta x$ and $\omega=2\tan(\omega_k\Delta t/2)/\Delta t$, where $\omega_k>0$ is the positive solution of the local numerical dispersion relation $4\sin^2(\omega_k \Delta t/2)/\Delta t^2=4\sin^2(k \Delta x/2)/\Delta x^2$. In the discrete version of $k_xE_y=\omega B_z$, it is necessary that we take $E_y=E_{s+j/2}^{n+1/2}$, and center $B_z$ on time-like faces $B_{r-k/2-i/2}^{n+1/2}=(B_{r-k/2}^{n}\!+\!B_{r-k/2-i}^{n}\!+\!B_{r-k/2}^{n+1}\!+\!B_{r-k/2-i}^{n+1})/4$. A similar relation holds for the $E_z$ and $B_y$ components, which are subdominant in our simulations. Using these momentum-space Faraday's law, the $k$-spectrum of right-propagating \textit{EM} waves ($k>0$) and left-propagating \textit{EM} waves ($k<0$) can be separated from the spatial Fourier transforms of electric and magnetic fields.

Results presented in Figs.~\ref{fig:part}-\ref{fig:total} are averaged over an ensemble of 200 simulations with statistically equivalent initial conditions. The ensemble average starts to show convergence for tens of realizations. In these simulations, temporal gauge $A^0=0$ is used, and periodic boundary conditions are employed for both $\phi_v$ and $A_e$. We choose resolutions $mdx=0.04$ and $mdt=0.005$, high enough that the fastest dynamics is resolved and the simulation results converge. The 1D box is large enough such that the laser does not transit the spatial domain before we terminate the simulations.

\section{\label{sec:conclusion}Discussion and Summary}
In this paper, we develop an algorithm for solving the Klein-Gordon-Maxwell's equations [Eqs.~(\ref{eq:KGMp}) and (\ref{eq:KGMA})], which can be used to model bosonic plasmas in the relativistic-quantum regime. This algorithm is derived by first discretizing the action [Eq.~(\ref{eq:Sd})] in a way that respects the U(1)-gauge symmetry. We then take variations with respect to the discrete fields to find their classical equations of motion. The resultant variational algorithm guarantees that the Bianchi identities, namely, $\nabla\cdot\mathbf{B}=0$ and the Faraday's law, are automatically and exactly satisfied. The remaining equations of motions are the discrete Gauss's law [Eq.~(\ref{eq:Ed})], which can be used to initialize the simulation; the discrete Klein-Gordon's equation [Eq.~(\ref{eq:phid})], which can be used to advance the charged field; and the discrete Maxwell-Amp\`ere's law [Eq.~(\ref{eq:Bd})], which can be used to advance the gauge field. After fixing a gauge, explicit scheme for advancing the discrete fields in time can be constructed (Appendix~\ref{appendix:B}, Fig.~\ref{fig:Schematics}). Our variational scheme respects local conservation laws (Appendix~\ref{appendix:A}), and can be easily parallelized using domain decomposition. Moreover, the numerical scheme we have developed can be inherently mimicked by quantum systems with local couplings \cite{Wiese13,Martinez16}, which can be efficiently realized using quantum parallelism \cite{Feynman86,Lloyd96}.

The numerical scheme we have developed has a number of advantages over conventional methods for simulating plasmas. As comparison, the two conventional methods that can fully simulate kinetic effects are the particle-in-cell (PIC) solvers and the Vlasov solvers. 
The PIC solvers represent particles in the continuum, while representing \textit{EM} fields on grids. Particles feel \textit{EM} fields through interpolations, and \textit{EM} fields feel particles through depositions. Using proper smoothing functions, these two steps can preserve gauge symmetry and symplectic structures, thereby respect local conservation properties when used in geometrical algorithms \cite{Squire12,Xiao13,Xiao15,Qin15}. Nevertheless, interpolation and deposition introduce artificial collisions, which are absent in physical systems.
In the other scheme, the Vlasov solvers, \textit{EM} fields are represented on the three-dimensional space, while particles are represented in the six-dimensional phase space. Particles are directly forced by fields on spatial grids, while the fields feel particles though velocity space integrals, which requires resolving three extra dimensions with substantial computational cost. 
In contrast, our algorithm represents both particles and \textit{EM} fields on the same grid. Therefore, there is no need for interpolations and depositions as in the case of PIC solvers, nor is there need for resolving extra velocity space dimensions as in the case of Vlasov solvers. 
Our algorithm folds the phase space dynamics of charged particles into the complex plane, and enables modeling of relativistic and quantum dynamics in regimes that cannot be described using semiclassical treatments. In the example of linear waves (Sec.~\ref{sec:examples}.\ref{sec:wave}), we show that relativistic-quantum wave dispersion relations can be recovered. Moreover, using the example of a gamma-ray laser interacting with a dense plasma slab (Sec.~\ref{sec:examples}.\ref{sec:pair}), we show that our algorithm naturally allows pair production when the laser intensity exceeds the Schwinger threshold.

Of course, the advantages of our algorithm come at an expense. The expense comes from the necessity of resolving the relativistic-quantum scales, which are much smaller than scales that classical plasma physics typically deals with. The coarsest resolution needed in relativistic quantum plasma simulations is determined by the lowest energy scale of the problem, which is the rest mass of electrons $\sim0.5$ MeV, corresponding to time scale of $\sim10^{-21}$ s, and spatial scale of $\sim 10^{-13}$ m. 
This resolution requirement can be seen from the discrete KG equation [Eq.~(\ref{eq:phin})], in which we must have $m\Delta t\ll 1$ in order for $\delta\phi_v\ll\phi_v$. Moreover, since we are solving a system of hyperbolic partial differential equations, 
the Courant--Friedrichs--Lewy (CFL) condition $\Delta t<\Delta x$ must be satisfied. 
Finally, it is worth noting that high resolution is required for large gauge fields. Since the gauge field appears through the Wilson's lines in complex exponentials [Eqs.~(\ref{eq:D0}) and (\ref{eq:D1})], the discrete theory is invariant under the gauge transformation $A_e\rightarrow A_e+2\pi/(q\Delta)$. Therefore, the discrete gauge field $A_e$ lives on the torus $\mathbb{T}^{1,3}$, which has a very different topology than $\mathbb{R}^{1,3}$. 
Therefore, the step size must be small enough such that $qA\Delta <2\pi$, in order to avoid exciting topological modes that are absent in the continuous theory.

In summary, we develop a variational algorithm for solving the Klein-Gordon-Maxwell equations, which described tree-level scalar QED in the classical-statistical regime and may be solved using quantum computers. We demonstrate that remarkably rich physics are contained in solutions to classical field equations, which can be used to model high-density plasmas interacting with short-wavelength electromagnetic fields. Our work uses scalar QED as a toy model to make the case that real-time lattice QED is a powerful tool for plasma physics in the strong-field regime, where relativistic-quantum scales overlap with collective plasma scales. The applications of lattice spinor-QED to laboratory relevant plasma conditions remain to be demonstrated in the future.

\begin{acknowledgments} 
The authors are grateful to Qun Wang and Sebastian Meuren for helpful discussions. This rsearch is supported by NNSA Grant No.~DE-NA0002948 and DOE Research Grant No.~DEAC02-09CH11466. Jianyuan Xiao is supported by National Magnetic Confinement Fusion Energy Research Project (2015GB111003, 2014GB124005), National Natural Science Foundation of China (NSFC-11575185, 11575186, 11305171), JSPS-NRF-NSFC A3 Foresight Program (NSFC-11261140328), Chinese Scholar Council (201506340103), Key Research Program of Frontier Sciences CAS (QYZDB-SSW-SYS004), and the GeoAlgorithmic Plasma Simulator (GAPS) Project. 
\end{acknowledgments}

\begin{appendices}
\appendix
\renewcommand{\appendixname}{APPENDIX}

\section{\label{appendix:A}\MakeUppercase{Geometric identities and local conservation laws}}
When discretizing the gauge 1-form $A$ and calculating the curvature 2-form $F=dA$ in Sec. \ref{sec:algorithm}.\ref{sec:discrete}, geometric structures of discrete exterior calculus are respected. Consequently, the identity $d^2=0$ holds for the discrete exterior derivative. In components, this Bianchi identity can be written as $0=dF=(\partial_\sigma F_{\mu\nu}+\partial_\mu F_{\nu\sigma}+\partial_\nu F_{\sigma\mu})dx^{\mu}\wedge dx^{\nu}\wedge dx^{\sigma}/3!$. One nontrivial identity, corresponding to all indexes being spatial, is $\nabla
\cdot\mathbf{B}=0$. When discretized, this identity becomes
\begin{eqnarray}
\label{eq:B0}
\frac{1}{\Delta_l}\Big(B_{r+\frac{l}{2}}^{n}-B_{r-\frac{l}{2}}^{n}\Big)=0.
\end{eqnarray}
The other nontrivial identity, corresponding to two spatial indexes and one temporal index, is the Faraday's law $\partial_t\mathbf{B}=-\nabla\times\mathbf{E}$, whose discrete version is
\begin{eqnarray}
\label{eq:E0}
&&\frac{1}{\Delta t}\Big(B_{r-\frac{i}{2}}^{n+1}-B_{r-\frac{i}{2}}^{n}\Big)=\frac{\epsilon_{ijk}}{\Delta_k}\Big(E_{s+\frac{j}{2}+k}^{n+\frac{1}{2}}-E_{s+\frac{j}{2}}^{n+\frac{1}{2}}\Big).
\end{eqnarray}
The above finite difference equations are automatically satisfied in our algorithm by geometric constructions, in contrast to standard elecromagnetic algorithms, such as the Yee's algorithm \cite{Yee66}, in which the Faraday's law needs to be solved as a dynamical equation. 

\begin{figure}[t]
	\renewcommand{\figurename}{FIG.}
	\includegraphics[angle=0,width=0.5\textwidth]{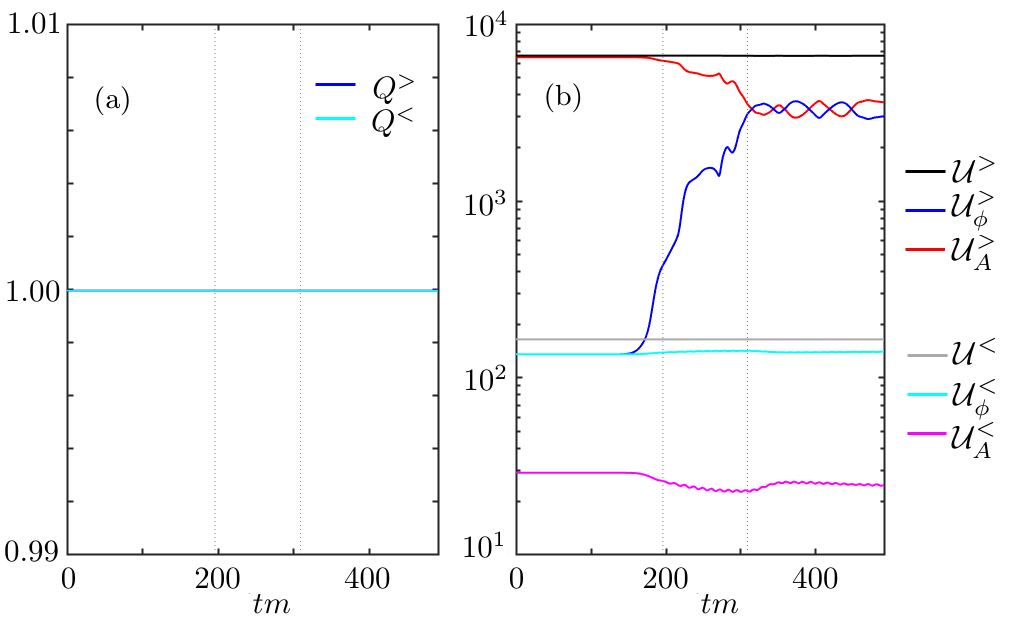}
	\caption{Evolution of total charge (a) and total energy (b) in the numerical example discussed in Sec.~\ref{sec:examples}.\ref{sec:pair}, where periodic boundary conditions are used. The total charge is constant in time up to the machine precision, both when $E<E_c$ (cyan) and $E>E_c$ (blue). When $E<E_c$ is below the Schwinger field, a small amount of energy is transfered from the electromagnetic field (magenta) to the charged field (cyan) due to wakefield acceleration and plasma wave excitation, while the total energy (gray) remains constant. In contrast, when $E>E_c$, a large amount of laser energy (red) is consumed by pair production. The energy of the charged field (blue) significantly increases, while the total energy (black) remains constant. The total charge $Q^{n+1/2}=\sum_s J_{s}^{n+1/2}$ is normalized by the total ion charge, and the total energy $\mathcal{U}^{n+1/2}=\sum_s\mathcal{H}_{s}^{n+1/2}$ is normalized by $m^3/\Delta x$. The vertical dashed gray lines mark the time when the laser pulse center enters and leaves the plasma plateau boundaries.}
	\label{fig:total}
\end{figure}

In addition to the above geometric identities, we also have a number of local conservation laws. The first is local charge conservation, which is a direct consequence of local U(1)-gauge symmetry. Under the continuous U(1)-gauge transformation
\begin{eqnarray}
\label{eq:U1phi}
\phi_{s}^{n}&\rightarrow&\phi_{s}^{n} e^{iq\alpha_{s}^{n}}, \\
A_{s}^{n+\frac{1}{2}}&\rightarrow& A_{s}^{n+\frac{1}{2}}+\frac{1}{\Delta t}(\alpha_{s}^{n+1}-\alpha_{s}^{n}),\\
\label{eq:GaugeAi}
A_{s+\frac{l}{2}}^{n}&\rightarrow& A_{s+\frac{l}{2}}^{n}+\frac{1}{\Delta_l}(\alpha_{s+l}^{n}-\alpha_{s}^{n}),
\end{eqnarray}
where $\alpha_{s}^n$ is any real-valued function living on vertexes. These transformations leave the discrete face-centered field strength tensor $F_f$ invariant, while transforming the pull-back covariant derivative by
\begin{eqnarray}
\label{eq:Cov}
(D_\mu^<\phi)_{s}^n\rightarrow e^{iq\alpha_{s}^{n}}(D_\mu^<\phi)_{s}^n.
\end{eqnarray}
Since the action $S_d$ is invariant, we can use the classical field equations $\delta S_d/\delta\phi_v=0$ and write
\begin{equation}
\frac{\delta S_d}{\delta \phi_{s}^{n}}\delta\phi_{s}^n+c.c.=0.
\end{equation}
Substituting in the infinitesimal transformation $\delta\phi_{s}^n=iq\alpha_{s}^n\phi_{s}^n$, the above identity is equivalent to the discrete charge conservation law
\begin{equation}
\label{eq:Jconserv}
\nonumber
\frac{1}{\Delta t} \Big(J_{s}^{n+\frac{1}{2}}-J_{s}^{n-\frac{1}{2}}\Big)=\frac{1}{\Delta_l} \Big(J_{s+\frac{l}{2}}^{n}-J_{s-\frac{l}{2}}^{n}\Big).
\end{equation}
Here, the charge density $J_{s}^{n+\frac{1}{2}}$ is given by Eq.~(\ref{eq:charge}), and the current density $J_{s+l/2}^{n}$ is given by Eq.~(\ref{eq:current}), and the sign is due to the Minkowski metric. It is straightforward to check that the above discrete charge conservation law is compatible with the discrete Gauss' law [Eq.~(\ref{eq:Ed})] and the discrete Maxwell-Amp\`ere's law [Eq.~(\ref{eq:Bd})].  In the numeric example discussed in Sec.~\ref{sec:examples}.\ref{sec:pair}, the total charge $Q^{n+1/2}=\sum_i J^{n+1/2}_i$ is constant up to the machine precision (Fig.~\ref{fig:total}a), both when the laser field is below ($Q^<$) and above ($Q^>$) the Schwinger field. 

Moreover, the discrete action $S_d$ is invariant under translations on the discrete spacetime manifold. 
Although the symmetry group in this case is discrete and hence the Noether's theorem does not immediately apply, we do have local energy conservation laws for the charged field and \textit{EM} fields separately when their coupling vanishes. Using the classical field equations $\delta S_d/\delta\phi_v=0$ and $\delta S_d/\delta A_{s+l/2}^n=0$, as well as the Bianchi identity, we have the following identity 
\begin{eqnarray}
\nonumber
0&=&\frac{\delta S_d}{\delta \phi_{s}^{n}}(\mathcal{D}_0\phi)_{s}^{n}+\frac{\delta S_d}{\delta \bar{\phi}_{s}^{n}}(\overline{\mathcal{D}_0\phi})_{s}^{n}\\
&+&\frac{\delta S_d}{\delta A_{s+l/2}^{n}}\frac{1}{2}\Big(E_{s+l/2}^{n+1/2}+E_{s+l/2}^{n-1/2}\Big)\\
\nonumber
&+&B_{r-l/2}^{n}\frac{1}{2}\Big[(d^2A)_{r-l/2}^{n+1/2}+(d^2A)_{r-l/2}^{n-1/2}\Big],
\end{eqnarray}
where the vertex-centered time covariant derivative 
\begin{equation}
(\mathcal{D}_0\phi)_{s}^{n}=\frac{e^{-iq\Delta t A_s^{n+\frac{1}{2}}}\phi_s^{n+1}-e^{iq\Delta t A_s^{n-\frac{1}{2}}}\phi_s^{n-1}}{2\Delta t}.
\end{equation}
After rearranging terms, the above identity gives rise to the local energy conservation law
\begin{eqnarray}
\label{eq:PEnergy}
\frac{\mathcal{H}_{s}^{n+1/2}\!-\!\mathcal{H}_{s}^{n-1/2}}{\Delta t}
=\frac{\mathcal{P}_{s+l/2}^{n}\!-\!\mathcal{P}_{s-l/2}^{n}}{\Delta_l}+\mathcal{O}(q\Delta^2), \hspace{15pt}
\end{eqnarray}
where the sign is again due to the Minkowski metric. The energy density can be separated into three terms
\begin{equation}
\mathcal{H}_{s}^{n+1/2}=\mathcal{H}_{s}^{n+1/2}[\phi]+\mathcal{H}_{s}^{n+1/2}[A]+h_{s}^{n+1/2},
\end{equation}
where the energy density of the charged field is
\begin{eqnarray}
\label{eq:Hphi}
\nonumber
\hspace{-2pt}\mathcal{H}_{s}^{n+\frac{1}{2}}[\phi] \!&=&\!\frac{1}{2}\Big[(D_0^<\phi)_s^{n+\frac{1}{2}}\!(\overline{D_0^<\phi})_s^{n+\frac{1}{2}}
\!+\!m^2\phi_s^{n}e^{iq\Delta tA_{s}^{n+\frac{1}{2}}}\bar{\phi}_s^{n+1}\\
&&\hspace{4pt}+(D_l^<\phi)_{s+\frac{l}{2}}^{n}e^{iq\Delta tA_{s}^{n+\frac{1}{2}}}(\overline{D_l^<\phi})_{s+\frac{l}{2}}^{n+1}\Big]\!+\!\text{c.c.},
\end{eqnarray}
and the energy density of the \textit{EM} fields is
\begin{eqnarray}
\label{eq:HA}
\mathcal{H}_{s}^{n+\frac{1}{2}}[A]=\frac{1}{2}\Big[\big(E_{s+\frac{l}{2}}^{n+\frac{1}{2}}\big)^2+B_{r-\frac{l}{2}}^{n+1} B_{r-\frac{l}{2}}^{n}\Big].
\end{eqnarray}
The energy density correction $h=\mathcal{O}(q\Delta^2)$ can take many different forms, each has a corresponding error term at finite-resolution. 
As expected, the energy density is U(1)-gauge invariant, so is the momentum density, which can be split into two terms 
\begin{equation}
\mathcal{P}_{s+l/2}^{n}=\mathcal{P}_{s+l/2}^{n}[\phi]+\mathcal{P}_{s+l/2}^{n}[A].
\end{equation}
The momentum density of the charged field is
\begin{equation}
\mathcal{P}_{s+\frac{l}{2}}^{n}[\phi]=(D_l^<\phi)_{s+\frac{l}{2}}^{n} e^{iq\Delta_lA_{s+\frac{l}{2}}^{n}}(\overline{\mathcal{D}_0^<\phi})_{s+l}^{n}+\text{c.c.},
\end{equation}
and the momentum density of the \textit{EM} fields $\mathcal{P}_i=-\mathcal{P}^i=-(\mathbf{E}\times\mathbf{B})^i$ is
\begin{eqnarray}
	\mathcal{P}_{s+\frac{i}{2}}^{n}[A]\!=\epsilon_{ijk}B_{r-\frac{j}{2}}^{n}\frac{1}{2}\Big(E_{s+i+\frac{k}{2}}^{n+\frac{1}{2}}\!+\!E_{s+i+\frac{k}{2}}^{n-\frac{1}{2}}\Big).
\end{eqnarray}
Since the stress-energy tensor $\mathcal{T}^{\mu\nu}$ is not a 2-form, neither the energy density $\mathcal{H}$ nor the momentum density $\mathcal{P}$ is well-defined on the discrete spacetime manifold. Hence, it can be shown, by enumerating combinations of U(1)-gauge invariant basis terms, that the resulting error in the local energy conservation law [Eq.~(\ref{eq:PEnergy})] is always second order. Except when the coupling $q\rightarrow0$, in which case the conservation law becomes exact even at finite spacetime resolutions. This remarkable feature would be lost if we had used the Wilsonian plaquettes in the discrete action instead. In examples discussed in Sec.~\ref{sec:examples}.\ref{sec:pair}, the total energy $\mathcal{U}^{n+1/2}\!=\!\sum_i\! \mathcal{H}^{n+1/2}_i$, whose error 
is of order $\mathcal{O}(qn\Delta t^2)$, is redistributed among $\phi$ and $A$ (Fig.~\ref{fig:total}b). The total energy fluctuates up to 6 ppm and $0.2 \%$ when the laser wakefield is below ($\mathcal{U}^<$) and above ($\mathcal{U}^>$) the Schwinger field.

\section{\label{appendix:B}\MakeUppercase{Numerical scheme}}
In this appendix, we list the four equations that are necessary for implementing the algorithm. We rewrite these equations from Sec.~\ref{sec:algorithm}.\ref{sec:EOM}, such that the explicit nature of the algorithm becomes apparent. The first step in the simulation is initializing values of $\phi^0_s, \phi^1_s, A^0_{s+l/2}$, and $A^{1/2}_s$ for all spatial indexes. This step is crucial and determines what physical system will be evolved subsequently.

The second step is calculating $A^1_{s+l/2}$ using the discrete Gauss' law [Eq.~(\ref{eq:Ed})], which can be rewritten as
\begin{eqnarray}
\label{eq:Gauss}
\frac{A^1_{s+l/2}-A^1_{s-l/2}}{\Delta_l}&=&\frac{A^0_{s+l/2}-A^0_{s-l/2}}{\Delta_l}+\Delta t J_s^{1/2}\\
\nonumber
&+&\frac{\Delta t}{\Delta l^2}\Big(A^{1/2}_{s+l}-2A^{1/2}_s+A^{1/2}_{s-l}\Big),
\end{eqnarray}
where all terms on the RHS are known. Since the LHS couples only two adjacent $A^1_{s+l/2}$ in each direction, the discrete Gauss' law is easier to solve than the Poisson's equation, which couples three nearest neighbors in each direction.

The third step is advancing the time-component of the gauge field $(A_s^{n-1/2},A_{s+l/2}^{n})\rightarrow A_s^{n+1/2}$. This step depends on the choice of the gauge condition. For example, when Lorenz gauge is used
\begin{eqnarray}
\label{eq:Lorenz}
A_s^{n+1/2}=A_s^{n-1/2}+C_l\Big(A^n_{s+l/2}-A^n_{s-l/2}\Big),
\end{eqnarray} 
where $C_l=\Delta t/\Delta_l$ is the dimensionless Courant number.
In comparison, when temporal gauge is used instead, $A^{n+1/2}_s=0$ and the time advance is trivial. Using the temporal gauge, one only needs to store values of $A_e$ at integer time steps $t=n$. However, when a background electric field is present, $A_{s+l/2}^{n}$ will grow indefinitely in the temporal gauge. In this case, long-time dynamics may be more accurately computed using the Lorenz gauge. 

In the fourth step, we can use the discrete KG equation [Eq.~(\ref{eq:phid})] to time advance the charged field $(\phi_s^{n-1}, \phi_s^{n};  A_{s+l/2}^{n}, A_s^{n\pm1/2})\rightarrow\phi_s^{n+1}$. The explicit time advance is given by
\begin{eqnarray}
\label{eq:phin}
\phi_s^{n+1}\!&=&\Big[(2-2C_l^2-\Delta t^2m^2)\phi_{s}^{n} -\phi_{s}^{n-1}e^{iq\Delta tA_{s}^{n-\frac{1}{2}}}\\
\nonumber
&+&C_l^2\Big(\phi_{s+l}^{n}e^{-iq\Delta_lA_{s+\frac{l}{2}}^{n}}+\phi_{s-l}^{n}e^{iq\Delta_lA_{s-\frac{l}{2}}^{n}}\Big)\Big]\!e^{iq\Delta tA_{s}^{n+\frac{1}{2}}}.
\end{eqnarray}
For free $\phi$ field, suppose the fluctuation is of the form $\exp(ip_lx^l-iEt)$, then the numerical dispersion relation of the massive particle is
\begin{eqnarray}
\nonumber
\frac{4}{\Delta t^2}\sin^2\frac{E \Delta t}{2}=\frac{4}{\Delta_l^2}\sin^2\frac{p_l \Delta_l}{2}+m^2,
\end{eqnarray}
which is consistent with the continuum energy-momentum relation $E^2=\mathbf{p}^2+m^2$ for relativistic particles. For the numerical solution to be stable, $E$ must be real, which holds only if for each $l=i,j,k$, the CFL condition $C_l<1$ is satisfied.

Finally, without relying on the fourth step, we can use the discrete Maxwell-Amp\`ere's law [Eq.~(\ref{eq:Bd})], simultaneously with the KG equation, to advance the spatial component of the gauge field $(A_{s+l/2}^{n-1},A_s^{n\pm1/2},A_{s+l/2}^{n};\phi_s^{n})\rightarrow A_{s+l/2}^{n+1}$. The explicit time advance is given by
\begin{eqnarray}
\label{eq:An}
\nonumber
A_{s+\frac{i}{2}}^{n+1}&=&A_{s+\frac{i}{2}}^{n}+C_i\Big(A_{s+i}^{n+\frac{1}{2}}-A_s^{n+\frac{1}{2}}\Big)\!+\!\Delta t^2J_{s+\frac{i}{2}}^n\\
&+&\!\Delta t\Big[E_{s+\frac{i}{2}}^{n-\frac{1}{2}}+\epsilon_{ijk}C_j\Big(B_{r-\frac{k}{2}}^n-B_{r-\frac{k}{2}-j}^n\Big)\Big],
\end{eqnarray}
where the RHS is known. For free gauge field, it is straight forward to show that the numerical solution is stable if and only if the CFL condition $C_l<1$ is satisfied.

\end{appendices}

%

\end{document}